\documentclass[12pt]{iopart}
\usepackage{latexsym,amsmath}
\usepackage{epsfig}
\newcommand{\C}{I\!\!\!\!C}   
\newcommand{\R}{I\!\!R}
\newcommand{\N}{I\!\!N}

\newcommand{\re}{{\rm Re}}
\newcommand{\im}{{\rm Im}}
\newcommand{\QED}{\quad\mbox{\rule[-4pt]{6pt}{6pt}}\vskip3mm\noindent}

\newcommand{\nin}{\mbox{$\hskip 6pt /\hskip -10pt\in\,$}}
\newcommand{\cf}{{\em c.f.$\;$\/}}
\newcommand{\ie}{{\em i.e.$\;$\/}}
\newcommand{\eg}{{\em e.g.$\;$\/}}

\newcommand{\proof}{{\em Proof\/:}}

\newcommand{\sgn}{{\rm sgn\,}}

\newcommand{\lra}{\longrightarrow}

\newcommand{\ra}{\rightarrow}

\newtheorem{claim}{Claim}
\newtheorem{theo}[claim]{Theorem}
\newtheorem{prop}[claim]{Proposition}

\newtheorem{lem}[claim]{Lemma}

\DeclareMathOperator{\Ai}{Ai}
\DeclareMathOperator{\Bi}{Bi}
\begin{document}
\title{Resonances In a Box
}
\author{George A. Hagedorn\dag\footnote[4]{Supported in part by the
National Science Foundation under grant number DMS--9703751.}
                and
                Bernhard Meller\ddag\footnote[3]{Research supported by
                FONDECYT Proyecto \#{}3970026} 
}
\address{\dag\ Department of Mathematics and Center for Statistical
Mechanics and Mathematical Physics,
Virginia Polytechnic Institute and State University,
Blacksburg, Virginia 24061--0123, U.S.A.}
\address{\ddag\ Facultad de F\'\i{}sica, P.U. Cat\'olica de Chile,
Casillo 306, Santiago 22, Chile}

\begin{abstract} We investigate a numerical method for studying resonances in
quantum mechanics. We prove rigorously that this method yields accurate
approximations to resonance energies and widths for shape resonances in the
semiclassical limit.
\end{abstract}
\maketitle
\section{Introduction}
In this paper we rigorously analyze the validity of a numerical technique for
studying resonances in quantum mechanics. The technique is called,
``A spherical box approach to resonances,'' by its inventors, Maier~\emph{et al}
\cite{MaiCedDom80a}. We prove that the technique yields correct
energies and lifetimes for shape resonances in the semiclassical limit.

The technique is an ``$L^{2}$ method,'' in contrast to
time--independent scattering theory methods, such as the calculation of phase
shifts near energies where a resonance is expected. These $L^2$ methods are
surveyed, \eg, in \cite{KukulinEtAl}.

The basic physical idea underlying all $L^{2}$ methods is that a resonance
wavefunction is a state that is concentrated mainly in the interaction
region. In contrast, states associated with the rest of the continuous
spectrum are not concentrated in any bounded interval.
As a consequence, when
the system is confined to a box that is large compared to the interaction
region and the size of the box is varied,
the resonance wavefunction is much less influenced than the states from
the rest of the continuous spectrum. This should be visible in the spectrum,
and is the basis of the technique we study.

To make this precise, we consider the Schr\"odinger operator
\begin{equation}\label{eq:Hamiltonian}
H:=D^{2} +V\,, \qquad D:= \frac{\hbar}{i}\frac{d}{dx}
\end{equation}
with a resonance producing potential \( V \) that is defined on all of $\R$.
We restrict the system to the interval $(-\ell,\,\ell)$ with
Dirichlet boundary conditions at \( x=\pm \ell \), and plot the eigenvalues of
the resulting operator $H(\ell)$ as a function of \( \ell \).

\begin{figure}[ht]
\begin{center}
\epsfig{figure=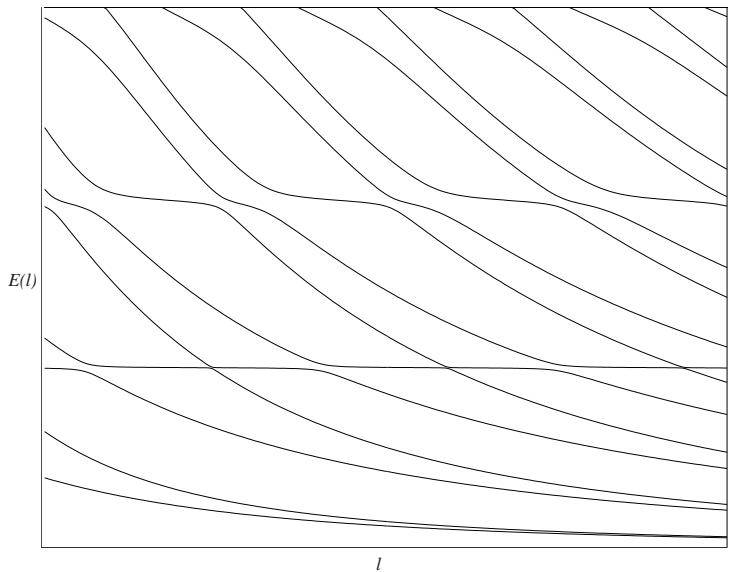,height=3in}
\caption{An example of the dependence of the eigenvalues on the box size $\ell$.}
\label{fig:lines}
\end{center}
\end{figure}

\begin{figure}[ht]
\begin{center}
\epsfig{figure=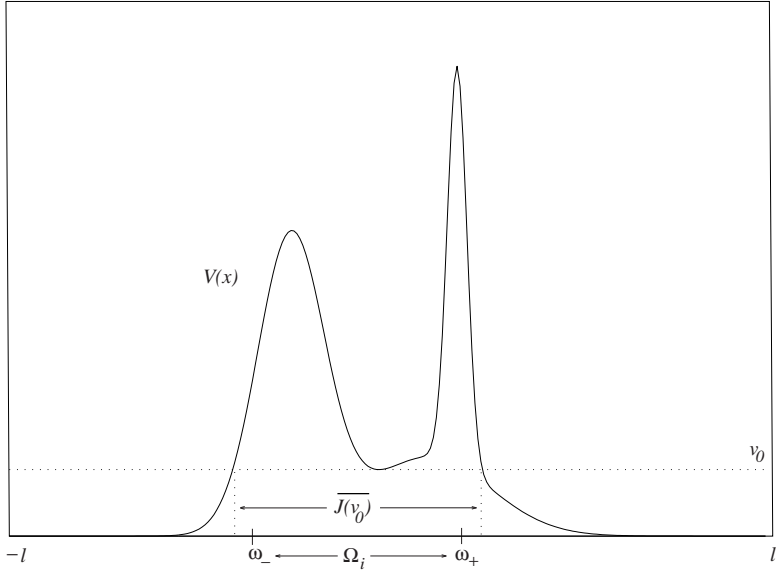,height=3in}
\end{center}
\caption{The potential associated to Figure~1 and relevant parameters}
\label{fig:pot}
\end{figure}

Figure~\ref{fig:lines} presents the results obtained by doing this for the
potential \( V \) that is depicted in Figure~\ref{fig:pot}.

In this figure, one can clearly distinguish between
eigenvalues that depend strongly on $\ell$ and others that seem to be
almost independent of $\ell$. Furthermore, there are avoided eigenvalue
crossings when a strongly dependent eigenvalue is close to an eigenvalue that
is almost independent of $\ell$. Note that in our example, eigenvalues are not
expected to cross \cite{WigVNeu27a}, since the potential has no apparent
symmetry properties.

In addition to relating the almost constant eigenvalues to resonance energies,
Maier~\emph{et al} \cite{MaiCedDom80a} also relate the sizes of the gaps in the
avoided crossings to the imaginary part (or width, or inverse lifetime) of the
resonance. In \cite{MaiCedDom80a}, spherically symmetric potentials are treated.
After the reduction to an angular momentum subspace, the particle can escape
to infinity in only one way, by increasing the radial coordinate $r$. In the
model we consider, the particle can escape toward either plus or minus infinity.
Since the probablilities for going in the two directions can be different, we
observe two different size gaps for each given resonance. This is obviously the
case in Figure~\ref{fig:lines}. For our model, the resonance width is related to
the larger of the two gaps.

In this paper, we provide rigorous justification of these results
in the semiclassical limit.
As a first step, we adopt a standard definition of a resonance that is presented
in \cite{AguCom71,Reed-SimonIV}. This definition identifies a
resonance with a complex eigenvalue of a suitably constructed analytic
family of operators obtained from the original Hamiltonian
(\ref{eq:Hamiltonian}).

In many instances, as in the case of shape resonances,
such a complex eigenvalue can be viewed as arising from the perturbation of
an eigenvalue embedded in the continuous spectrum.
We take this viewpoint and employ
the framework of, ``The Shape Resonance,'' \cite{ComDucKleSei87a} by
Combes \emph{et al}. We temporarily impose supplementary Dirichlet boundary
conditions at points $\omega_{\pm}$ to decouple the interaction region from the
rest of $\R$. This yields an unperturbed operator on all of $\R$ that has
embedded eigenvalues whose eigenfunctions are supported in the interaction
region. Removal of these Dirichlet conditions perturbs the embedded
eigenvalues to produce the resonances (that are realized as complex eigenvalues
of certain non-self-adjoint operators). The perturbation calculations are
facilitated by the use of Krein's formula \cite{ComDucKleSei87a}.

To relate the resonances of $H$ defined on $L^2(\R)$ to the almost
$\ell$--independent eigenvalues of $H(\ell)$, we show that the techniques of
\cite{ComDucKleSei87a} can also be applied in a box to study $H(\ell)$.
We then employ the following strategy:
For small values of $\hbar$, resonances of $H$ are very close to embedded
eigenvalues of $H$ with supplementary Dirichlet conditions at $\omega_{\pm}$.
For $\ell >\max\,\{|\omega_+|,\,|\omega_-|\}$, these embedded eigenvalues are
also eigenvalues of $H(\ell)$ with supplementary Dirichlet conditions at
$\omega_{\pm}$. For large $\ell$ and small $\hbar$, removal of these
supplementary Dirichlet conditions perturbs these eigenvalues only slightly.
Thus, the resonances of $H$ are near eigenvalues of $H(\ell)$.
These results are made precise in Theorem~\ref{theo:1}.

This approach also allows us to prove rigorously that the gap in the
avoided crossing is on the order of the square root of the resonance width,
in accordance with \cite{MaiCedDom80a}. We accomplish this by relating both the
gap and the resonance width to the thickness of the potential barrier as
measured by the Agmon distance \cite{Agmon}. The relationship between resonance widths and
Agmon distances is already established in \cite{ComDucKleSei87a}, so we need
only examine the relationship between the Agmon distances and the gaps in the
avoided crossings. This is done in Theorem~\ref{theo:2}.

\section{Hypothesis and Results}
For simplicity, we assume the potential $V$ to be bounded.
We wish to study resonances that are produced by a single well and to avoid
asymptotically degenerate eigenvalues with an exponentially small separation in
$\hbar$.
Furthermore, we want the bottom \( v_{0} \) of the well to be above the
scattering threshold.
We force this situation by imposing a hypothesis
that can be expressed nicely with the help of the notion of
\textsl{the classical forbidden region at energy $E$}. This is defined as
                \[J(E):=\{x\in\R:\ V(x)>E\}\,.\]
Our precise hypothesis is the following:
\begin{itemize}
        \item[\textbf{(H1)}] $V\in C^{1}(\R)$ is bounded and has a
        local minimum $v_{0}$ at \( x_{0}\), such that
        $\overline{J(v_{0})}$ is connected, and
        $\limsup_{|x|\ra\infty}V(x)<v_{0}$.
\end{itemize}
By translating the origin if necessary, we choose an interior region
\[
\Omega_{i}:=(\omega_{-},\,\omega_{+}),\quad\mbox{ with }
\omega_{-}<0\mbox{ and }\omega_{+}>0,\mbox{ such that }\quad
\overline{\Omega}_{i}\setminus\{x_{0}\}\subset J(v_{0}).\,
\]
We define the exterior region to be
$\Omega_{e}:=\R\setminus\overline{\Omega}_{i}$, and let
$\Omega_{e}^-=(-\infty,\,\omega_-)$ and
$\Omega_{e}^+=(\omega_+,\,\infty)$.
We define the decoupled comparison operator \( H^{d} \) as having the same
symbol as \( H \), but with
supplementary Dirichlet conditions at $\omega_-$ and $\omega_+$.
This operator decomposes into
\[
H^{d}=H^{i}\oplus H^{e}
\quad\mbox{with}\quad {\cal D}(H^{\alpha})=
{\cal H}^{1}_{0}\cap{\cal H}^{2}(\Omega_{\alpha}),\quad
\mbox{where}\quad\alpha\in\{i,\,e\}\,.
\]
\noindent
Since we want to focus on shape resonances, we impose a hypothesis that prevents
resonances from being produced in the exterior region for energies near
\( v_{0} \). We phrase this hypothesis in terms of a non-trapping condition
\cite{BriComDuc87a}:
We say \textsl{the potential $V$ is non-trapping in $\Omega_{e}$ at
energy $E$ (abbreviated $E$ is NT)}, if the following condition is
satisfied for \( \alpha\in\{-,\,+\} \):
          \begin{equation}
   \exists S>0,\ \forall x\in\Omega_{e}^{\alpha}\setminus J(E), \quad
   \frac{x\!-\!\omega_{\alpha}}{x}\big( 2(V(x)\!-\!E)+xV'(x)\big)<-S
   \,.
   \label{eq:VirialCDKS}
           \end{equation}
We assume:
\begin{itemize}
        \item[\textbf{(H2)}] $v_{0}$ is NT.
\end{itemize}
Note that formula (\ref{eq:VirialCDKS}) is equivalent to the more standard
virial condition
        \[ \exists \tilde{S}>0,\ \forall x\in\Omega_{e}\setminus J(E), \quad
         2(V(x)-E)+xV'(x)<-\tilde{S}\,.
                 \]
Furthermore (\ref{eq:VirialCDKS}) implies the ``exterior'' virial condition
\[ \exists S>0,\ \forall x\in\Omega_{e}^{\alpha}\setminus J(E), \quad
                        \big( 2(V(x)-E)+(x-\omega_{\alpha})V'(x)\big)<-S,\
                        \alpha\in\{-,\,+\}\,.
\]
Our third hypothesis concerns analyticity under exterior dilation.
For $\theta\in\R$, we define
\begin{xalignat*}{2}
U_\theta : L^2(\R) &\ra L^2(\R)\quad\mbox{by} &&\\
U_\theta : \phi &\mapsto \sqrt{r_\theta'}\,\phi\circ r_\theta\quad 
\mbox{where}&
r_\theta(x) &:= \left\{
\begin{array}{cl} \omega_-+e^\theta(x-\omega_-),& x<\omega_-\\
x,& x\in(\omega_-,\,\omega_+) .\\
\omega_+ +e^\theta(x-\omega_+),& x>\omega_+
\end{array}\right.
\end{xalignat*}
We then assume:
\begin{itemize}
        \item[\textbf{(H3)}] $V_\theta:=U_\theta V U_\theta^{-1}$
        defined initially for $\theta\in\R$, has an analytic continuation as
        a bounded operator to the strip
$\{ \theta\in\C : |\im\,\theta | < \beta\}$, for some $\beta\in (0,\,\pi/4)$.
\end{itemize}
For $\theta\in\R$ we also define the operators
\begin{math}
  H_\theta := U_\theta H U_\theta^{-1}
\end{math}
and
\(H^d_\theta := U_\theta H^d U_\theta^{-1} \).
It is a straightforward calculation to obtain the associated symbol
\begin{align*}
U_\theta (D^2+V)U_\theta^{-1} =\, &{r_\theta '}^{-2} D^{2}+V\circ
r_\theta, \\
\mbox{where\hspace{4cm}}
[&{r_\theta '}^{-2} D^{2}u](x)=\left\{
\begin{aligned}
-\hbar^2 u''(x),&\quad x\in(\omega_-,\,\omega_+)\\
-\hbar^2 e^{-2\theta} u''(x),&\quad x\nin[\omega_-,\,\omega_+]\,.
\end{aligned}
\right.
\end{align*}
Since $U_\theta$ is a unitary operator on $L^2(\R)$ for $\theta\in\R$,
we easily compute the domains for the operators
\(H^d_\theta \)
and  \begin{math}H_\theta\end{math},
for $\theta\in \R$:
\begin{align}\notag
 {\cal D}(H^d_\theta)&={\cal D}(H^i)\oplus {\cal D}(H^e),\\
 \label{eq:Htheta}
 {\cal D}(H_\theta) &=
 \{u_i\oplus u_e\in {\cal H}^2(\Omega_i)\oplus {\cal H}^2(\Omega_e) :
   u_e(\omega_\pm)\!=\! e^{\frac{\theta}{2}}u_i(\omega_\pm),\,
   u_e'(\omega_\pm)\!=\! e^{\frac{3\theta}{2}}u_i'(\omega_\pm)\}.
\end{align}
We define the restrictions of these operators to the box $(-\ell,\,\ell)$
to be
\begin{align}\notag
H^d_\theta(\ell)&:={r_\theta '}^{-2} D^{2}+V\circ r_\theta
\quad\mbox{on}\quad {\cal D}(H^d_\theta) \cap {\cal H}^1_0((-\ell,\ell)),
\quad\mbox{and}\\
H_\theta(\ell)&:={r_\theta '}^{-2} D^{2}+V\circ r_\theta
\quad\mbox{on}\quad {\cal D}(H_\theta) \cap {\cal H}^1_0((-\ell,\ell))\,.
\label{eq:Hthetaell}
\end{align}
For $\theta=0$, $H_{\theta=0}(\ell)$ is simply the Schr\"odinger operator
$H(\ell)$ described in the introduction that is used to produce plots, such as
Figure~1.

The following lemma describes the analytic continuations of these families
of operators to complex values of $\theta$:
\begin{lem} Hypotheses \textbf{(H1)}--\textbf{(H3)} imply the following two
conclusions:
\begin{enumerate}
\item $\{H^d_{\theta},\, |\im\theta|<\beta\}$ and
$\{H^d_{\theta}(\ell),\, |\im\theta|<\beta\}$ are self-adjoint analytic
families of Type (A) of m-sectorial operators.
\item $\{H_{\theta},\, |\im\theta|<\beta\}$ and
$\{H_{\theta}(\ell),\, |\im\theta|<\beta\}$ are self-adjoint analytic
families of operators.
\end{enumerate}
\end{lem}
\proof\/  These conclusions for the families $H^d_\theta$
and $H_\theta$ are proved in \cite{ComDucKleSei87a}. The same proofs apply for
the families $H^d_\theta(\ell)$ and $H_\theta(\ell)$ since the proofs in
\cite{ComDucKleSei87a} make no use of
the (un)boundedness of $\Omega_e$.\QED

We next recall the Agmon distance \cite{Agmon}, that we denote by the
symbol \( d_{E} \). It is the distance associated to the pseudo-metric
\( ds^{2}:=\max\{0,V(x)-E\}dx^{2}\,\).
We introduce the abbreviations
\begin{xalignat*}{3}
        d_{v_{0}}^{\alpha}&:=
        d_{v_{0}}(x_{0},\,\alpha\ell),\ \,\alpha\in\{-,\,+\},&
                &\mbox{and}&
        d^{\star}&:= \min\{d_{v_{0}}^{-},\,d_{v_{0}}^{+}\}
\,.
\end{xalignat*}

The following theorem gives precise information about the resonance on the line
and the ``resonance in the box.'' Its first conclusion follows from
\cite{ComDucKleSei87a}.

\begin{theo}\label{theo:1}
        Assume \textbf{(H1)}--\textbf{(H3)}
        and that $E^{d}$ is the $n^{th}$ eigenvalue of $H^{i}$.
\begin{itemize}
\item[(i)] For any \( \vartheta\in(0,1) \) and sufficiently small $\hbar$, there
        exists $\beta_0\in(0,\beta)$, such that $H_{i\beta_0}$
        has a (complex) eigenvalue $E$ close to \( E^{d} \) that satisfies:
        \begin{equation*}\label{eq:tunnel exp}
             E=E^{d}+\sum_{n\geq 1}\frac{t^{n}\sigma_{n}}{n!},
            \quad\mbox{with}\quad
            t=o(e^{-2\vartheta d^{\star}/\hbar}),\quad \mbox{where}\quad
            \sigma_{n}=o(1),\ \forall n\geq 1.
        \end{equation*}
\item[(ii)] The same is true for the operator $H_{i\beta_0}(\ell)$.
        Furthermore its eigenvalue is stable in the sense of Kato
        \textrm{\cite[Sec. VIII.1.4]{Kato}}, as the box size $\ell$ tends to
        infinity. As $\ell$ tends to infinity, this eigenvalue
        converges to the corresponding eigenvalue of $H_{i\beta_0}$.
\item[(iii)] For sufficiently small \( \hbar \) and those values of
        \( \ell \), for which there exist positive constants \(c\)
        and \( N \), such that
        \(\mbox{\em dist}(E^{d}, \sigma (H^{d}(\ell))\setminus \{E^{d}\})
        \geq c\hbar^{N}\),
        there exists a real eigenvalue of \( H(\ell) \) close to \( E^{d} \)
        that satisfies the same type of expansion as above.
\end{itemize}
\end{theo}
\textbf{Remark:}
(a) Note that in \cite{ComDucKleSei87a},
the theorem is stated with \( d^{\star} \) replaced
by \( d_{v_{0}}(x_{0},\partial\Omega_{i})  \). Due to the possible choices
of \( \omega_{\pm} \), the difference between the two quantities can be
made arbitrarily small and can be absorbed into \( \vartheta \).
But then, how small \( \hbar \) must be chosen, depends on \( \vartheta \).

(b) In the third conclusion of this theorem, one cannot expect
uniform results in \(\ell\) and \( \hbar \).
The eigenvalues of the exterior operator $H^e(\ell)$
have different dependence on
\(\ell\) and \( \hbar \)
than the eigenvalues of the interior operator $H^i$.
The condition
{dist}\((E^{d}, \sigma (H^{d}(\ell))\setminus \{E^{d}\}) \geq c\hbar^{N}\)
is technical; we cannot handle exponentially closely spaced eigenvalues.
It is well known that under our hypotheses, the eigenvalues of $H^i$
near the bottom of the well (close to $v_0$) cannot be spaced more closely than
${\cal O}(\hbar^\gamma)$. Here, the constant $\gamma$ is strictly
smaller than 2. Its value depends on how flat the bottom of the well is.
In order to prove that eigenvalues from $H^e(\ell)$ do not cause
{dist}\((E^{d}, \sigma (H^{d}(\ell))\setminus \{E^{d}\}) \geq c\hbar^{N}\)
to be violated for all $\ell$, we would need
an additional assumption on the decay of the potential.
For example, together with dilation analyticity, it
would be enough to assume that $V$ tends to a limit at infinity like
$|x|^{-\epsilon}$ for any $\epsilon>0$.
\vskip3mm\noindent
We now turn our attention to the gaps in the avoided crossings that occur
in graphs of the eigenvalues of $H(\ell)$.
For this part of our analysis, we replace hypotheses \textbf{(H2)} and
\textbf{(H3)} by the following:
\begin{itemize}
\item[\textbf{(H4)}] $V\in C^3(\R)$. For
    $x\in\Omega_e\setminus \overline{J(v_0)}$, the potential obeys $V(x)< v_0$
    and there exist two constants $v_\pm<v_0$,
    such that $V-v_\pm = {\cal O}(|x|^{-\epsilon})$ as
    $x$ tends to $\pm\infty$. Furthermore for $n=1,2$, we have
    $V^{(n)}= {\cal O}(|x|^{-\epsilon-1})$ as $x$ tends to $\pm\infty$.
\end{itemize}
This hypothesis allows us to use WKB estimates to analyze the behavior of
eigenvalues of $H^e(\ell)$. We note that  $H^e(\ell)$ decomposes into the direct
sum of \( H^{e}_{-}(\ell) \) and \( H^{e}_{+}(\ell) \), where
\( H^{e}_{-}(\ell) \) acts on \( L^{2}((-\ell,\,\omega_{-})) \)
and \( H^{e}_{+}(\ell) \) acts on \( L^{2}((\omega_{+},\,\ell)) \).

We have the following result on the gaps:
\begin{theo}\label{theo:2}
Assume \textbf{(H1)} and \textbf{(H4)}.
Suppose \( E^{d}\) is an eigenvalue of \( H^i \) and of $H^e_\alpha(\ell_0)$,
but not of $H^e_{-\alpha}(\ell_0)$. Assume it satisfies
    \(\mbox{\em dist}(E^{d}, \sigma (H^{e}_{-\alpha}(\ell_0)))
    \geq c\hbar^{N}\),
for some positive constants \( c\) and \( N\) and $\alpha\in\{-,\,+\}$.
Then we have the following:
For fixed values of \( \hbar \) that are sufficiently small,
there exists a neighborhood \( {\cal V}(\ell_{0}) \)
of \( \ell_{0} \), such that for all $\ell$ in \( {\cal V}(\ell_{0}) \),
\( H(\ell) \) has two eigenvalues $E_+$ and $E_{-}$ that are
exponentially close to \( E^{d} \).
These two eigenvalues are separated by a gap that satisfies
\begin{equation*}
    \min_{\ell\in{\cal V}(\ell_{0})}\{ |E_{+}-E_{-}|\}
        =\left|\sum_{n\geq 1}\frac{(t_{1}+t_{2})^{n}\sigma_{n}}{n!}
        \right|\,,
            \text{ where}\quad \sigma_{n}=o(1),\
        \forall n\geq 1.
\end{equation*}
In this estimate, $t_1$ and $t_2$ satisfy the following for any
\(\vartheta\in(0,1) \):
 \begin{alignat*}{3} 
             &t_{1}&=t_{2}=o(\exp({-\vartheta d^{-}_{v_0}/\hbar}))\quad
             &\text{if}\quad&
             E^{d}&\in\sigma(H^{i})\cap\sigma(H^{e}_{-}(\ell_{0}))\,,
                         \\
                         &t_{1}&=t_{2}=o(\exp({-\vartheta d^{+}_{v_0}/\hbar}))\quad
             &\text{if}\quad&
             E^{d}&\in\sigma(H^{i})\cap\sigma(H^{e}_{+}(\ell_{0}))\,.
  \end{alignat*}
\end{theo}
\textbf{Remark:} (a) Here, \( \hbar \) does not depend on
\( \vartheta \).

(b) The width of the resonance is given by the tunneling
parameter $t$ according to Theorem~\ref{theo:1}. We do not
know whether the resonance is going to tunnel to the left or right, so we
only obtain the estimate
\( t= o(\exp({-\vartheta 2 d^{\star}/\hbar}))\).
In Theorem~\ref{theo:2} we know to which side the resonance escapes, and
the result is more precise. We also obtain estimates for both of the
gap sizes that can occur in the avoided crossings for a given resonance.
Since $d^{\star}=\min\{d_{v_{0}}^{-},\,d_{v_{0}}^{+}\}$, Theorem~\ref{theo:2}
shows that the larger gap is of the same order as the square root of the
resonance width.
We again note that in
\cite{MaiCedDom80a}, a radial symmetric situation is studied, so that
there is only one way for the resonance to escape, and hence only one gap size.

(c) The eigenvalues of \( H^{i} \) are obviously independent of \( \ell \),
but not of \( \omega_{\pm} \). Thus, it might seem that the condition
of having a double eigenvalue
is crucially dependent on the choice of \( \omega_{\pm} \).
This is not case:
From Theorem~\ref{theo:1}~(iii) we
see that the eigenvalues of \( H^{i} \) vary only by an exponentially
small quantity in \( \hbar \) when the \( \omega_{\pm} \) are varied.
For the eigenvalues of \( H^{e}_{\pm}(\ell) \), we show in
\ref{app:2}, that \textbf{(H4)} implies
that eigenvalues \( E\in \sigma(H^{e}_{\alpha}(\ell))\) that belong to an
interval \( (v_{0},v_{0}+\delta) \) are related to \( \hbar \), \( \ell \), and
a quantum number \( m \) by the asymptotic formula
\[
E=v_{\alpha}+\left((m+\frac{3}{4})\frac{\pi\hbar}{\ell}\right)^{2}\left(1+
{\cal O}(\hbar) +{\cal O}(\ell^{-\epsilon})\right)\,,
\quad\alpha\in\{-,\,+\}
\,.
\]
We thus have the following consequence:
Suppose, for example, that the \(n\)-th eigenvalue \( E^{d} \) of \( H^{i} \)
coincides with an eigenvalue of \( H^{e}_{+}(\ell_{0}) \) for some choice of
\( \omega_{\pm} \), and that \( E^{d} \) is at least a distance
of \( {\cal O}(\hbar^{N}) \) from the spectrum of \( H^{e}_{-}(\ell_0) \).
Then for any other choice of \( \omega_{\pm} \), there exists an \( \ell \) in
a neighborhood of \( \ell_{0} \), such that \( E^{d} \) is
an eigenvalue of \( H^{e}_{+}(\ell) \),
and the distance from \( E^{d} \) to the spectrum of \( H^{e}_{-}(\ell) \)
is still at least \( {\cal O}(\hbar^{N}) \).
\section{The Proofs}
Inspection of the proofs of \cite{ComDucKleSei87a} for
Theorem~\ref{theo:1}~(i) shows that they are valid whether or not
$\Omega_e$ is bounded. Furthermore, these proofs can be separated into two
parts: The first is a formal algebraic part that shows the stability
of the eigenvalue of \( H^{i} \) for the whole operator and constructs
the asymptotic expansion of the perturbed eigenvalue in powers
of the tunneling parameter \( t \).
It is quite simple and short.
The second part is the justification of these algebraic formulas with the
corresponding estimates.
This part is more complicated and involves estimation of the operators
involved in Krein's formula.

We present the formal algebraic part, which is needed in all of
the situations treated in Theorems~\ref{theo:1}~(ii), \ref{theo:1}~(iii), and
Theorem~\ref{theo:2}.  We do this in Section~\ref{sec:S} in the
context of Theorem~\ref{theo:1}~(iii).
In Section~\ref{sec:S2}, we treat the stability of the resonance eigenvalue
of $H_{i\beta_0}(\ell)$ as the box size $\ell$ tends to infinity.
Finally, in Section~\ref{sec:S3} we prove Theorem~\ref{theo:2}. In the
Appendix, we recall Krein's formula and present the more technical estimates,
including the WKB estimates.

We omit the estimates required to prove the existence and the
series expansion of the eigenvalue of $H_{i\beta_0}(\ell)$ because they are
identical to those in \cite{ComDucKleSei87a}.

\subsection{Stability and Tunneling Expansion for The Box}
\label{sec:S}
We view \( H(\ell) \) as a perturbation of \( H^{d}(\ell) \).
This perturbation involves two Dirichlet conditions. It is most easily
approached by way of Krein's formula, that exhibits the
difference of the resolvents of \( H(\ell) \) and \( H^{d}(\ell) \)
as a rank two operator.

The norm of this rank two operator is not small. However,
because the Dirichlet conditions are imposed inside the classically
forbidden region, its norm does not explode in proportion to the inverse of
the distance from the spectrum to the spectral parameter in the resolvents.
This allows us to choose the parameters in such a way that the
resolvent of the resolvent of \( H^{d}(\ell) \) is small in norm, and
we can still use perturbation theory.

The tunneling expansion is based upon a Feshbach type reduction of the
eigenvalue equation with respect to the unperturbed
eigenprojection. This leads to an implicit equation that we solve by using the
Lagrange inversion formula.
\subsubsection{Stability}
To simplify the notation, we suppress the $\ell$ dependence in many of the
formulas. We define
\[
R^{d}(z):=(H^{d}(\ell)-z)^{-1}\quad\mbox{and}
\quad R(z):= (H(\ell)-z)^{-1}\,.
\]
We choose a contour $\Gamma$ that lies in the resolvent set of
\( H^{d}(\ell) \) and encloses only \( E^{d} \) in \( \sigma(H^{d}(\ell)) \).
We then choose a point $z_0$ in the intersection of the resolvent sets
of \( H(\ell) \) and \( H^{d}(\ell) \), but outside of $\Gamma$.
By using the identity
\begin{equation}
        \left(R^{d}(z_{0})-\frac{1}{z-z_{0}}\right)^{-1}
= -(z-z_{0})-(z-z_{0})^{2}R^{d}(z),
        \label{eq:identity}
\end{equation}
we obtain the following expression for the eigenprojection \(
P^{d}\equiv P^{d}(\ell) \)
associated to \( E^{d} \):
\[ P^{d}\,=\,-\,\frac{1}{2\pi i}\,\int_{\Gamma}\,R^{d}(z)\,dz
    \,=\,-\,\frac{1}{2\pi i}\,\int_{\widetilde{\Gamma}}\,
         \left(R^{d}(z_{0})-\widetilde{z}\right)^{-1}\!d\widetilde{z}\,,
\]
where $\{\,\widetilde{\Gamma}:=\widetilde{z}\in\C:\,
\widetilde{z} =\frac{1}{z-z_{0}},\,z\in\Gamma\,\}$.
By defining
\[ \pi(z_0):= (H(\ell)-z_0)^{-1}-(H^{d}(\ell)-z_{0})^{-1}\,,
\]
we can formally write the eigenprojection \( P\equiv P(\ell) \) associated to the
perturbed eigenvalue \( E \) as
\[ P \,=\,-\,\frac{1}{2\pi i}\,\int_{\widetilde{\Gamma}}\,
         \left(R^{d}(z_{0})-\widetilde{z}\right)^{-1}
\left(1+\pi(z_{0})\left(R^{d}(z_{0})-
         \widetilde{z}\right)^{-1}\right)^{-1}
         d\widetilde{z}\,.
\]
If we can choose $\Gamma $ and $z_0$, such that
\(\left\|\,\pi(z_{0})\left(R^{d}(z_{0})
    -\widetilde{z}\right)^{-1}\,\right\|<1\),
then the inverse term in the integral for $P$ can be computed by
geometric series. Then the eigenprojection is well defined, and
by standard arguments,
we can deduce the stability of the eigenvalue for \( H(\ell) \).

To see that we can do this, fix any \( n\in\N \). Let
\begin{equation}\label{eq:defi delta}
\Delta:= \mbox{dist}
(E^{d}, \sigma (H^{d}(\ell))\setminus \{E^{d}\}),
\quad \mbox{and fix}\quad
r\in[\min\{\hbar^{n},\,\frac{1}{2}\Delta\},\,\frac{1}{2}\Delta ]\,.
\end{equation}
Note that by hypothesis, \( \Delta\geq c\hbar^{N} \), for some \( N\in\N \),
and that we can choose \( r \) to be as small as any power of \( \hbar \).

We define \( \Gamma:=\{z\in\C:|z-E^{d}| =r\} \)
and \( z_{0}= E^{d}+2ir \). Then, formula (\ref{eq:identity}) implies
\( \left(R^{d}(z_{0})-\widetilde{z}\right)^{-1}={\cal O}(r) \).
Thus, the stability follows from the following proposition that we prove in
\ref{app:1.2}:
\begin{prop}\label{prop:pi}
         \( \pi(z_{0})={\cal O}(1)  \).
\end{prop}
\subsubsection{Tunneling expansion}
\label{sec:TE}
Since we have proven the stability of the eigenvalue and constructed $P(\ell)$,
we can write the eigenvalue equation as
\[ R(z_{0})P(\ell)= \frac{1}{E-z_{0}}P(\ell)\,.\]
We perform a Feshbach type reduction to this equation, with respect to the
projections \( P^{d} \) and \( Q^{d}=1-P^{d} \).
We define the ``reduced'' resolvent
\[
\widehat{R}(z;\,z_{0}):=Q^{d}\,
{\left(Q^{d}(R(z_{0})-z)Q^{d}\right)}^{-1}\,Q^{d}
\,. \]
It satisfies the following estimate:
%
\begin{prop}\label{prop:hatR}
        For any \( z \) in the disc delimited by \( \Gamma \), one has
        \( \widehat{R}(\frac{1}{z-z_{0}};\,z_{0}) ={\cal O}(r) \).
\end{prop}
\proof\/ If we replace the \( R(z_{0}) \) by
\( R^{d}(z_{0}) \) in the definition of
\( \widehat{R}(z;\,z_{0}) \), we obtain a trivial result. The conclusion
to the proposition is obtained by applying perturbation theory to this trivial
result.\QED
For \( ({E\!-\!z_{0}})^{-1} \)
the reduction yields the implicit equation
\[ (\frac{1}{E\!-\!z_{0}}-\frac{1}{E^{d}\!-\!z_{0}})P^{d}P =
P^{d}\left(\pi(z_{0}) -\pi(z_{0})\,
\widehat{R}(\frac{1}{E\!-\!z_{0}};\,z_{0})\, \pi(z_{0}) \right)P^{d}P\,. \]
Using properties of the trace and the factorization
\( \pi(z)=\hbar A^{\star}(\bar z)B(z) \), \cf \ref{app:1.2}, we obtain
\begin{align}\notag
        \frac{1}{E\!-\!z_{0}} -\frac{1}{E^{d}\!-\!z_{0}}\,&=\,
        \hbar\,\mbox{Tr} \left(B(z_{0}) P^{d} A^{\star}(z_{0})
        \Big(1-\hbar B(z_{0})\,\widehat{R}(\frac{1}{E\!-\!z_{0}};\,z_{0})
           \,A^{\star}(\bar{z}_{0})\Big)\right)\,,\\
\intertext{or equivalently}
\label{eq:implicit F}
        \frac{1}{E\!-\!z_{0}} -\frac{1}{E^{d}\!-\!z_{0}}\,&=\,
        t\, s\,(\frac{1}{E\!-\!z_{0}})\,,
\end{align}
where (suppressing \( z_{0}\) in \( A \) and \( B \))
\begin{equation}
        t:=\hbar\,|\mbox{Tr}(B P^{d}A^{\star})|\quad\mbox{and}\quad
        s(z):=  \frac{1}{t}\,\mbox{Tr} \left(\hbar B
        P^{d}A^{\star}(1-\hbar B\widehat{R}(z;z_{0})A^{\star})\right)\,.
        \label{eq:tands}
\end{equation}
For any $z$ in the disc delimited by $\Gamma$ and $\tilde{z}=\frac{1}{z-z_{0}}$,
we have the following estimate on \( s(\tilde{z}) \):
\[
|s(\tilde{z})|\,\leq\,\|1-\hbar B\widehat{R}(\tilde{z};z_{0})A^{\star}\|
\,=\,1+{\cal O}(r).
\]
This follows from Proposition~\ref{prop:hatR} and the bound on \( \pi \),
\cf \ref{app:1.2}.
If we can establish the estimate \( t=o(e^{-2\vartheta d^{\star}/\hbar}) \)
of Theorem~\ref{theo:1}, then equation \eqref{eq:implicit F}
can be solved with Lagrange's inversion formula
\cite[p.250]{Dieudonne}
\[ \frac{1}{E\!-\!z_{0}}=\frac{1}{E^{d}\!-\!z_{0}}+\sum_{n\geq
1}\frac{t^{n}}{n!}\left[\frac{d^{n-1}}{dz^{n-1}}s^{n}
\right](\frac{1}{E^{d}\!-\!z_{0}})
=:\frac{1}{E^{d}\!-\!z_{0}}+\sum_{n\geq
1}\frac{t^{n}}{n!}\widetilde{\sigma}_{n}
\,.\]
Multiplying by \( (E\!-\!z_{0})(E^{d}\!-\!z_{0}) \) and rearranging, we obtain
\[
 E = E^{d}- (E\!-\!z_{0})(E^{d}\!-\!z_{0})
 \sum_{n\geq 1}\frac{t^{n}}{n!}\widetilde{\sigma}_{n}
 = E^{d}-\sum_{k\geq 1} (z_{0}\!-\!E^{d})^{k+1}
 \biggl(\sum_{n\geq
 1}\frac{t^{n}}{n!}\widetilde{\sigma}_{n}\biggr)^{k}\,.
\]
We estimate the coefficients \( \widetilde{\sigma}_{n} \) by using
the Cauchy formula
\[ \widetilde{\sigma}_{n} = \frac{(n-1)!}{2\pi i}\,
\int_{\widetilde{\Gamma}}\,
\frac{s(\widetilde{z})^{n}}{(\frac{1}{E^{d}\!-\!z_{0}}-\widetilde{z})^{n}}
\,d\widetilde{z},\quad\text{and}\quad
(\frac{1}{E^{d}\!-\!z_{0}}-\widetilde{z})^{-1}= {\cal O}(r)\,.
\]
We define
\( \sigma_{n}:= (E\!-\!z_{0})(E^{d}\!-\!z_{0})\widetilde{\sigma}_{n} \)
and easily obtain the estimate \( \sigma_{n}=o(1) \) of Theorem~\ref{theo:1}.
\subsubsection{The tunneling parameter}
The above calculation relies on
the estimate \( t=o(e^{-2\vartheta d^{\star}/\hbar}) \). To prove this, we
note that if \( \phi_{d} \) denotes the eigenfunction associated to \(
E^{d} \), then using the definitions and estimations of \ref{app:1.2},
\begin{align*}
          t\,&\leq\,
        \hbar\,\|B\phi_{d}\|\,\|A\phi_{d}\|
        \,\leq\,\hbar^{2}\,\|TRT^{\star}\|\, \|B\phi_{d}\|^{2} \\
&=\,\frac{\hbar^{2}\,\|TRT^{\star}\|}{|E^{d}\!-\!z_{0}|^{2}}\,
\|T^{d}D\phi_{d}\|^{2}
        \,\leq\,\frac{c\hbar^{3}}{4r^{2}}\,
         (|\phi_{d}'(\omega_{-})|^{2}+|\phi_{d}'(\omega_{+})|^{2})\,.
\end{align*}
For each part of Theorem~\ref{theo:1}, we can estimate the expression
\(|\phi_{d}'(\omega_{-})|^{2}+|\phi_{d}'(\omega_{+})|^{2}\)
by the well known decay estimates of Agmon \cite{Agmon}.
This implies the results of Theorem~\ref{theo:1}.
\subsection{Stability as The Box Size Tends to Infinity}
\label{sec:S2}
We consider the operator
\begin{equation}
        H^{D}_{\theta}(\ell):= H_{\theta}(\ell)\oplus
        H^{ee}_{\theta}(\ell)\,,
        \label{eq:HD}
\end{equation}
where \( H_{\theta}(\ell) \) is the operator
defined in \eqref{eq:Hthetaell}, and
\[
H^{ee}_{\theta}(\ell):= e^{-2\theta}D^{2}+ V\circ r_{\theta}
\quad\mbox{on} \quad
{\cal H}^{1}_{0}\cap{\cal H}^{2} (\R\setminus [-\ell,\ell])
\,.
\]
It is easy to see that \( H^{ee}_{\theta}(\ell) \) is an analytic
family of Type (A) in \( \theta \), and that we have the following
resolvent estimate:
\begin{prop}\label{prop:Hee}
Assume \textbf{(H1)}--\textbf{(H3)} and let $S$ denote the
constant in the non-trapping condition \textbf{(H2)}.
Let $\nu=\{z\in\C : |\re\,z -v_0|< S/4,\,\im\,z > -S/4\}$.
Then
\[ \forall\, z\in\nu,
\qquad \|R^{ee}_{i\beta}(z)\|\,\leq\,
\frac{4}{|\beta | S}\,(1+{\cal O}(\beta)).\]
\end{prop}
\proof\/  \( H^{ee}_{i\beta}(\ell) \) decomposes into a direct sum of operators
that act on $L^2((-\infty,\,-\ell))$ and $L^2((\ell,\,\infty))$.
We consider only the term associated to the interval \( (\ell,\infty) \);
analogous formulas hold for the other term. We mimic arguments of
\cite{BriComDuc87a}.
For \( u\in{\cal H}^{1}_{0}\cap{\cal H}^{2} ((\ell,\infty))\)
and any \( v\in L^{2}((\ell,\infty)) \), we have
\begin{align*}
        \|v\|\,\|(H^{ee}_{i\beta}(\ell)-z)u\| &
        \,\geq\,\re\,((H^{ee}_{i\beta}(\ell)-z)u,\,v)\,.
\end{align*}
For \( \beta>0 \) we use this with \( v= -ie^{-i2\beta}u\) to obtain
\begin{align*}
        \re\,((H^{ee}_{i\beta}(\ell)-z)u,v)\,&=\,
        -\,\im\,(e^{2i\beta}(e^{-2i\beta}D^{2}+V\circ r_{i\beta}-z)u,\,u)\\
        &=\,-\,\im\,(e^{2i\beta}(V\circ r_{i\beta}-z)u,\,u)\\
        &=\,-\,((\beta\,( 2(V-\re\,z)+(x\!-\!\omega_{+})V' -\im\,z ) +
        {\cal O}(\beta^{2}))u,\,u)\\
        &>\,(\beta\,(S-2(v_{0}\!-\! \re\,z)+\im\,z)+{\cal O}(\beta^{2}))\,
        \|u\|^{2} \\
&>\,\left(\,\frac{\beta\,S}{4} +{\cal O}(\beta^{2})\,\right)\,
        \|u\|^{2}\,.
\end{align*}
For negative \( \beta \) we repeat this calculation with
\( v\!=\! ie^{-i2\beta}u \). This proves the proposition.\QED
\noindent
We now fix \( \theta=i\beta_{0} \) as in \cite{ComDucKleSei87a}. With the
definitions of \( z_{0}\) and \( \widetilde{\Gamma} \) as in
Section~\ref{sec:S}, we define
\[ P_{i\beta_0}(\ell) \,=\,-\,\frac{1}{2\pi i}\,\int_{\widetilde{\Gamma}}\,
         \left((H_{i\beta_0}(\ell)\!-\!z_{0})^{-1}-\widetilde{z}\right)^{-1}
                 \oplus
                 \left((H^{ee}_{i\beta_0}(\ell)\!-\!z_{0})^{-1}-\widetilde{z}\right)^{-1}
         d\widetilde{z}\,.
\]
Here, \( P_{i\beta_0}(\ell) \) projects onto the eigenspace for the
eigenvalue \( E \in \sigma( H_{i\beta_0}(\ell)) \), but does so in the space
\( L^{2}(\R) \).
To prove stability of the eigenvalue in the generalized sense (\cf Kato,
\cite[Sec. VIII.1.4]{Kato}), we must show
that \(  P_{i\beta_0}(\ell)\stackrel{s}{\lra} P_{i\beta_0} \)
as \( \ell \) tends to \( \infty \), where
\[ P_{i\beta_0} \,=\,-\,\frac{1}{2\pi i}\,\int_{\widetilde{\Gamma}}\,
         \left((H_{i\beta_0}\!-\!z_{0})^{-1}-\widetilde{z}\right)^{-1}
                 d\widetilde{z}\,.
\]
It is shown in \cite{ComDucKleSei87a} that for sufficiently small \( \hbar \),
\( \left((H_{i\beta_0}\!-\!z_{0})^{-1}-\widetilde{z}\right)^{-1} =
{\cal O}(r) \),
uniformly on \( \widetilde{\Gamma} \).
The estimates of \cite{ComDucKleSei87a} are also valid for
\( \left((H_{i\beta_0}(\ell)\!-\!z_{0})^{-1}-\widetilde{z}\right)^{-1} \).
So, from Proposition~\ref{prop:Hee} and identity \eqref{eq:identity}, we see
that,
\[ \left((H^{D}_{i\beta_0}(\ell)\!-\!z_{0})^{-1}-\widetilde{z}\right)^{-1} =
{\cal O}(r)\,,
\]
uniformly on \( \widetilde{\Gamma} \).
Thus, we need only show that for any \( u\in L^{2}(\R) \),
\[
\lim_{\ell\ra\infty}\left\|\left( (H_{i\beta_0}-z_{0})^{-1}-
(H^{D}_{i\beta_0}(\ell)-z_{0})^{-1}\right) u\right\| =0\,,
\]
uniformly in \( \hbar \).
This is shown in \ref{app:1.1}
\subsection{Proof of Theorem 3}
\label{sec:S3}
In the degenerate case, we must solve for two eigenvalues. So, we cannot
\emph{a priori} use the Lagrange inversion formula to solve equation
\eqref{eq:implicit F} in the disc delimited by \( \Gamma \).

However, we could use the formula if one of the solutions were known to
be \( \frac{1}{E^{d}\!-\!z_{0}} \). This would happen if \( \pi \) were a rank
one operator. In that case, the spectra of \( H^{d} \) and \( H \)
would intertwine, and as a consequence,
at the crossing of two eigenvalues of \( H^{d} \) there would have to be an
eigenvalue of \( H \).

In our situation such a scenario can be realized by lifting the two
Dirichlet conditions one after the other.

It suffices to consider the case where
\( E^{d}\in\sigma(H^{i})\cap\sigma(H^{e}_{+}(\ell_{0}))\).
In the first step, we consider the operators
\[        H^{d}_{-}(\ell) := H^{e}_{-}(\ell)\oplus H^{i}
        \quad\text{and}\quad
        H_{-}(\ell) := D^{2} +V \quad\text{on}\quad
        L^{2}((-\ell,\omega_{+}))\,.
\]
By hypothesis, \( \hbar \) is small and fixed, and \( H^{i} \) has the
eigenvalue \( E^{d} \), which for \( \ell=\ell_{0} \) is a distance
of \( {\cal O}(\hbar^{N}) \) from the rest of the spectrum of
\( H^{d}_{-}(\ell_{0}) \), \ie \( E^{d} \) is a simple, conveniently
isolated  eigenvalue of \( H^{d}_{-}(\ell_{0}) \).
Thus, the analog Theorem~\ref{theo:1}~(iii) is valid:
\begin{lem}\label{lem:cutbox}
Assume the hypotheses of Theorem~\ref{theo:2}
with \( E^{d}\in\sigma(H^{i})\cap\sigma(H^{e}_{+}(\ell_{0}))\).
Then there exists a neighborhood of \( {\cal V}(\ell_{0}) \), of size
  \( c\hbar^{N} \), such that for each
  \( \ell\in{\cal V}(\ell_{0}) \), the operator
  \( H_{-}(\ell) \) has an eigenvalue \( E_{-} \) close to \( E^{d} \)
  that satisfies the following for any \(\vartheta\in(0,1) \)
  \begin{equation*}
     E_{-}=E^{d}+\sum_{n\geq 1}\frac{t^{n}\sigma_{n}}{n!}
         \quad\text{with}\quad
     t= o(e^{-2\vartheta d_{v_0}(\omega_{-},x_{0})/\hbar})
         \quad\text{and}\quad
     \sigma_{n}=o(1),\,\forall n\geq 1.
  \end{equation*}
\end{lem}
\proof\/ We first note that as we vary \( \ell \), with the restriction that
\( |\ell-\ell_{0}|\leq c\hbar^{N} \),
\( E^{d} \) remains isolated from the rest of the spectrum by a distance of
size \( c\hbar^{N} \).
Thus, we can prove the lemma by mimicking the proof of
Theorem~\ref{theo:1}~(iii).\QED
\noindent
For the second step, we note that due to the behavior of \( E^{e}_{+}(\ell) \)
there exists an \( \ell_{1}\in {\cal V}(\ell_{0}) \), such that
\[ E_{-}= E^{e}_{+}(\ell_{1}) \,.\]
We now use the intertwining of the spectra of
\( H_{-}(\ell_{1})\oplus H^{e}_{+}(\ell_{1}) \) and \( H(\ell_{1}) \).
We obtain the following lemma by using the techniques we used for
Lemma~\ref{lem:cutbox} and
noting that the eigenfunction \( \phi_{d} \) associated to \( E_{-} \)
has the form \( \phi_{d}=\phi_{-}\oplus\phi_{+} \), where
\( H_{-}(\ell_{1})\phi_{-}=E_{-}\phi_{-} \) and
\( H^{e}_{+}(\ell_{1})\phi_{+}=E_{-}\phi_{+} \):
\begin{lem} Assume \textbf{H(1)} and \textbf{(H4)} and that \( E_{-} \)
is a double eigenvalue of \(H_{-}(\ell_{1})\oplus
H^{e}_{+}(\ell_{1})\) as constructed above. Then the operator
\( H(\ell_{1}) \) has two  eigenvalues \(E_{-}\) and  \( E_{+} \)
that satisfy
\begin{equation*}
   E_{+}=E_{-}
   +\sum_{n\geq 1}\frac{(t_{1}+t_{2})^{n}\sigma_{n}}{n!}
      \quad\text{with}\quad
   \sigma_{n}=o(1),\ \forall n\geq 1\,,
\end{equation*}
where, for any \(\vartheta\in(0,1) \),
\begin{xalignat*}{3}
    t_{1}&=o(e^{-2\vartheta d_{v_{0}}(x_{0},\omega_{+})/\hbar})&
      &\mbox{and}&
    t_{2}&=o(e^{-2\vartheta d_{v_{0}}(\omega_{+},\ell_{1})/\hbar})\,.
        \end{xalignat*}
\end{lem}

The last step in the proof of Theorem~\ref{theo:2} is to note
that the first two steps can be done for any admissible \( \omega_{+} \).
The \( n \)-th eigenvalue \( E^{d} \) of \( H^{i} \) changes by only
an exponentially small amount in \( \hbar \) when \( \omega_{+} \) is
varied, so it remains properly isolated from
\( \sigma( H^{e}_{-}(\ell_{0})) \). Furthermore, by the behavior of the exterior
eigenvalues, there exists an \( \ell_{2} \) in a neighborhood of \( \ell_{0} \),
such that the new \( E_{-} \) is also an eigenvalue of
\( H^{e}_{+}(\ell_{2}) \).
The optimal estimate is obtained when \( t_{1}=t_{2} \), in which case
we have \( t_{1}=t_{2}= o(e^{-\vartheta d^{+}_{v_{0}}/\hbar})\).
\appendix
\section{Krein's Formula}
Since we need Krein's formula for one and two supplementary Dirichlet boundary
conditions, taken at different points depending on the situation, we wish to
present the formula in a general setting. On
the other hand, for simplicity, we leave out the exterior dilation.
We deal with this only when necessary.

Suppose \(n\geq 2\), and \( -\infty\leq x_{0}<x_{1}<\ldots<x_{n}\leq \infty \)
are specified. Let \( \Omega:=(x_{0},x_{n}) \) and
\( \Omega_{k}:=(x_{k-1},x_{k}) \) for \( k=1,\ldots,n \).
Let \( H:=D^{2}+V \) be a Schr\"odinger operator on \( \Omega \),
with self-adjoint boundary conditions at \( x_{0} \) and \( x_{n} \),
and let \( H^{d} \) be the corresponding decoupled operator with
supplementary Dirichlet conditions at \( x_{1},\,x_{2},\ldots ,\,x_{n-1} \).
Denote their resolvents by \( R \) and \( R^{d} \), respectively.

Let \( z\in \rho(H)\cap \rho(H^{d}) \) and \( u,v\in L^{2}(\R) \).
Define \( \hat{u}:=R^{d}(z)u \) and \( \hat{v}:=R({z})^{\star}v \).
Clearly, \( \hat{u}\in {\cal D}(H^{d})\), and thus,
\( \hat{u}=\oplus_{k=1}^{n}\hat{u}_{k} \) with
\( \hat{u}(x_{k})=0,\,k=1,\ldots,n-1\).
We have
\begin{align*}
   \left( (R(z)-R^{d}(z))u,v\right)\,&=\,
   (u,\hat{v})-(\hat u,v)
                \\
   &=\,\sum_{k=1}^{n}\,(D^{2}\hat{u}_{k},\hat{v})_{\Omega_{k}} -
           (\hat{u}_{k},D^{2}\hat{v})_{\Omega_{k}}\\
   &=\,- \hbar^{2}\,\sum_{k=1}^{n}\,
           \hat{u}'_{k}\,\overline{\hat{v}}\Big|_{\partial\Omega_{k}}
           \\ 
   &=\,\hbar^{2}\,\sum_{k=1}^{n-1}\,
           \left(\,\hat{u}'_{k+1}\!-\!\hat{u}'_{k}\,\right)\,
           \overline{\hat{v}}\big|_{x_{k}}\,.
\end{align*}
We use standard Sobolev space notation and define functionals
\( T^{j}_{x_{k}} \) by the following relations, where
\( f\in\oplus_{k=1}^{n}{\cal H}^{1}(\Omega_{k}) \):
\[
T^{j}_{x_{k}}:{\cal H}^{1}(\Omega_{j})\lra \C,\quad
T^{j}_{x_{k}}f:=\lim_{y\ra x,\,y\in\Omega_{j}} f(y),\quad\mbox{for}\quad
j=k,\,k+1,\quad k=1,\ldots,n-1.
\]
If \( T^{k}_{x_{k}}f= T^{k+1}_{x_{k}}f \) for all $f$, we simply write
\( T_{x_{k}} \).
It is well known that \( T^{j}_{x_{k}} \) is compact, and consequently,
\( (T^{j}_{x_{k}})^{\star}:\C^{2}\ra{\cal H}^{-1}(\Omega_{j}) \) is
continuous. Furthermore,
Lemma 4 of Section III of \cite{ComDucKleSei87a} shows that whenever
\( \chi\in C^{\infty}_{0}(\R) \) satisfies \( \chi(x_{k})= 1 \) for
\( k=1,\ldots,n-1 \),
\begin{equation}
\left\|T^{j}_{x_{k}}u\right\|^{2}\,\leq\,
2 \hbar^{-1}\left\|\chi u_{j}\right\|
\left\|D \chi u_{j}\right\|
\,\leq\,c\hbar^{-1}\left\|\chi u_{j}\right\|_{{\cal H}^{1}}\,,
\quad\mbox{for}\quad j =k,\,k+1.
\label{eq:trace}
\end{equation}
Finally, we define
\[
T^{-}:=\begin{pmatrix} T^{1}_{x_{1}} \\ \vdots\\ T^{n-1}_{x_{n-1}}\end{pmatrix},
\quad
T^{+}:=\begin{pmatrix} T^{2}_{x_{1}} \\ \vdots\\ T^{n}_{x_{n-1}}\end{pmatrix},
\quad
T^{d}:= -\,T^{-}\oplus T^{+}
\quad\mbox{and}\quad
T:=\begin{pmatrix} T_{x_{1}} \\ \vdots\\ T_{x_{n-1}}\end{pmatrix}.
\]
With these definitions, we have the following formula,
\[
\left((R(z)-R^{d}(z))u,\,v\right)
\,=\,\hbar\,\big(R(z) T^{\star}iT^{d}DR^{d}(z)u,\,v\big)\,,
\]
where all the multiplications are understood to be matrix multiplications.
\subsection{Applying Krein's Formula for Theorem~\ref{theo:1}~(iii)}
\label{app:1.2}
In the proof of Theorem~\ref{theo:1}~(iii), we have
$x_{0}=-\ell$, $x_{1}=-\omega_{-}$, $x_{2}=\omega_{+}$, $x_{3}=\ell$,
\[
R^{d}(z)=(H^{d}(\ell)-z)^{-1},\quad\mbox{and}
\quad R(z)= (H(\ell)-z)^{-1}\,.
\]
Following \cite{ComDucKleSei87a}, we define
\[
B(z):= i T^{d} D R^{d}(z) \quad\mbox{and}\quad
                        A(z):=  T R(z)\,.
\]
Since \( H \) is self-adjoint, we can write
\[
\pi(z)=R(z)-R^{d}(z)=\hbar\,A^{\star}(\overline{z}) B(z).
\]
Furthermore, since \( TR^{d}(z)=0 \), we have
\[
T \pi(z) = TR(z)= A(z)=
\hbar\,T R(z) T^{\star} B(z) \,.
\]
We combine the two formulas to obtain
\[ \pi(z)= \hbar^{2}B^{\star}(\overline{z}) \,
TR(z)T^{\star}\,B(z)\,.
\]
Proposition~\ref{prop:pi} now follows from
\begin{prop}\label{lem:esti B}
        Let \( z_{0}= E^{d}+2ir \). Fix any \( N\in \N \). Then for
        sufficiently small \( \hbar \) and any
        \( r\in[\min\{\hbar^{N},\,\frac{1}{2}\Delta\},\,\frac{1}{2}\Delta ]\),
        \[ B(z_{0})={\cal O}(\hbar^{-1/2})\qquad\mbox{and}\qquad
        TR(z_{0})T^{\star}={\cal O}(\hbar^{-1})\,. \]
\end{prop}
\emph{Proof:} The assertion on \( TR(z_{0})T^{\star} \) is proved in
step 5 of the proof of Theorem~III.3 of \cite{ComDucKleSei87a}.
As for \( B(z_{0}) \) we have
\[
\|B(z_{0})\|^{2}= \|T^{1}_{\omega_{-}}DR^{e}(z_{0})\|^{2}
+\|T^{2}_{\omega_{-}}DR^{i}(z_{0})\|^{2}
+\|T^{2}_{\omega_{+}}DR^{i}(z_{0})\|^{2}
+\|T^{3}_{\omega_{-}}DR^{e}(z_{0})\|^{2}\,,
\]
where \( R^{i}(z_{0}) := (H^{i}-z_{0})^{-1}\) and
\( R^{e}(z_{0}) := (H^{e}(\ell)-z_{0})^{-1}\). Let
\( \chi  \) be a \( C^{\infty}_{0} \) function supported around
\(\omega_{\pm}\) such that \( \chi(\omega_{\pm})=1 \).
Using the estimate
\eqref{eq:trace}, it suffices to find a uniform bound on the expressions
\[\chi DR^{i}(z_{0}),\quad D\chi DR^{i}(z_{0}),
\qquad\mbox{and}\qquad
\chi DR^{e}(z_{0}),\quad D\chi DR^{e}(z_{0})\,.
\]
We choose \( \chi \), such that
\( V(x)-v_{0}\geq\varepsilon>0 \) for \( x\in \mbox{supp}\,\chi \).
Then steps 1
and 2 of the proof of Theorem~III.3 of \cite{ComDucKleSei87a} show
that all theses terms are uniformly bounded.\QED
\subsection{Applying Krein's Formula for Theorem~\ref{theo:1}~(ii)}
\label{app:1.1}
Here we consider the difference of the resolvents of
the operators \( H_{i\beta_0} \) defined by
formula~\eqref{eq:Htheta} of
the introduction and \( H^{D}_{i\beta_0}(\ell) \) defined by
formula~\eqref{eq:HD}. In this case,
$n=3$, $x_{0}=-\infty$, $x_{1}=-\ell$, $x_{2}=\ell$, and $x_{3}=\infty $.
The difference of the resolvents is
\[   R_{i\beta_0}(z_{0})-R_{i\beta_0}^{d}(z_{0})
      = \hbar R_{i\beta_0}(z_{0}) T^{\star}
          ie^{-2i\beta_0}T^{d}DR^{d}_{i\beta_0}(z_{0})\,.
 \]
Let \( \chi  \) be a \( C^{\infty}_{0} \) function supported around
\(\pm\ell\), with \( \chi(\pm\ell)=1 \).
To show that \(T^{d} D R^{D}_{i\beta_0}(z_{0})\) and
\( T R_{-i\beta}(\overline{z}_{0}) \)
are uniformly bounded operators
we use the estimate \eqref{eq:trace}. Thus, it suffices to
show that \(\chi D R^{D}_{i\beta_0}(z_{0})\) and
\( \chi R_{-i\beta}(\overline{z}_{0}) \)
are uniformly bounded operators from \( L^{2} \) to \(  {\cal H}^{1}\).
If that is true, then for \( u\in L^{2}(\R) \), we have
\begin{align*}
        \left\|\left( R_{i\beta_0}(z_{0})-R_{i\beta_0}^{d}(z_{0})\right) u\right\|
        &\leq
        \left\|TR_{-i\beta}(\overline{z}_{0}) \right\|
        (|\hat{u}(-\ell)|+|\hat{u}(\ell)|) \\
        &\leq c(|\hat{u}(-\ell)|+|\hat{u}(\ell)|)
        \stackrel{\ell\ra\infty}{\lra}0 \,,
\end{align*}
uniformly in \( \hbar \), since
\( \hat u =\chi DR^{D}_{i\beta_0}(z_{0})u \in {\cal H}^{1}(\R) \).

We now address the required uniform bounds.
Commuting \( \chi \) and \( D \), we need only consider
\( \chi DR^{D}_{i\beta_0}(z_{0}) \) and \( \chi D^{2}R^{D}_{i\beta_0}(z_{0}) \).
The expressions for
\( T R_{-i\beta}(\overline{z}_{0}) \) are analogous and can be
treated the same way.
The formula
\[
\|\chi DR^{D}_{i\beta_0}(z_{0})\|^{2}=
\re\,R^{D}_{i\beta_0}(z_{0})^{\star} \chi^{2}
\left(D^{2}-2\hbar^{2}(\chi^{2})''\right) R^{D}_{i\beta_0}(z_{0})
\]
shows
that it is sufficient to bound \( \chi D^{2}R^{D}_{i\beta_0}(z_{0}) \) and
\( \chi R^{D}_{i\beta_0}(z_{0}) \).
We have
\begin{align*}
        \chi D^{2}R^{D}_{i\beta_0}(z_{0}) &=
         e^{2i\beta}\chi \left(H^{D}_{i\beta_0}-z_{0}-(V\circ r_{i\beta_0}-z_{0})
         \right)R^{D}_{i\beta_0}(z_{0})\\
         &=e^{2i\beta}\left(1-(V\circ r_{i\beta_0}-z_{0})
         \chi R^{D}_{i\beta_0}(z_{0})\right)
\end{align*}
For \( \chi R^{D}_{i\beta_0}(z_{0}) \), we set
\( \pi_{i\beta_0}(z_{0}):=(H_{i\beta_0}(\ell)\!-\!z)^{-1}
-(H^{d}_{i\beta_0}(\ell)\!-\!z)^{-1} \), and then write
\begin{align*}
        \chi R^{D}_{i\beta_0}(z_{0}) &=
         \chi \left(R^{i}(z_{0})\oplus (H^{e}_{i\beta_0}(\ell)-z_{0})^{-1}
         +\pi_{i\beta_0}(z_{0})
         \right) \oplus R^{ee}_{i\beta_0}(z_{0})\\
         &=\chi\left((H^{e}_{i\beta_0}(\ell)-z_{0})^{-1}
         +\pi_{i\beta_0}(z_{0})\right) \oplus
          R^{ee}_{i\beta_0}(z_{0})\,.
\end{align*}
The right hand side is uniformly bounded in \( \hbar \) and \( \ell \)
by Propostion~\ref{prop:Hee} and
Lemma~II.3  and Theorem~III.3
of \cite{ComDucKleSei87a}, which are also valid for
\( (H^{e}_{i\beta_0}(\ell)-z_{0})^{-1} \) and \( \pi_{i\beta_0}(z_{0})
\), respectively.

\section{WKB Estimates}
\label{app:2}
For these estimates, we follow Olver \cite[Ch. 11]{Olver}.
The goal is to find approximate solutions to the differential equation
\begin{equation}
         -\hbar^{2}w'' +(V-E)w=0\,
        \label{eq:Schroedinger}
\end{equation}
in \( \Omega_{e} \) with \( v_{0}+\delta>E>v_{0} \) for some
positive \( \delta \).
Due to either the non-trapping condition or the explicit assumption
\textbf{(H4)}, there is at most one turning point in each of the
intervals \( (\omega_{+},\,\infty) \) and \((-\infty,\,\omega_{-}) \).
There is exactly one, if \( \delta \) is sufficiently small.
We assume $\delta$ has been chosen so that this is the case.

It suffices to consider the interval \( (\omega_{+},\infty) \), and we
denote the turning point by \( x_{t} \).
We define a new independent variable \( \xi:= s(x) \) by
\[ s(x)\,s'(x)^{2}= E-V(x)\,, \quad s(x_{t})=0\,,\quad s'(x_{t})>0\,.
\]
By integration, we obtain
\[
\xi= \sgn(x-x_{t})\biggl(\frac{3}{2}{\cal S}(x)\biggr)^{2/3}
\quad\mbox{where}\quad
{\cal S}(x):= \int_{\min\{x,x_{t}\}}^{\max\{x,x_{t}\}}
\sqrt{|V(t)-E|}\,dt
\,.
\]
Note that \( \sgn( V(x)-E) = \sgn(x_{t}-x) \).
It is easy to check that under our conditions,
Theorem~3.1 of \cite[Ch.~11]{Olver} shows
that equation \eqref{eq:Schroedinger} has two \( C^{2} \) solutions
\( w_{1} \) and \( w_{2} \) in \( (\omega_{+},\infty) \), such that
\begin{equation}\label{eq:WKB1tp sol}
        \begin{split}
                 w_{1}(x;\hbar) & =
                 s'(x)^{-1/2}\biggl(
                 \Bi(-\xi/\hbar^{2/3})+{\cal O}(\hbar\Bi(-\xi/\hbar^{2/3}))\biggr)\,,\\
                 w_{2}(x;\hbar) & =
                 s'(x)^{-1/2}\biggl(
                 \Ai(-\xi/\hbar^{2/3})+{\cal O}(\hbar\Ai(-\xi/\hbar^{2/3}))\biggr)\,.
        \end{split}
\end{equation}
Higher order approximations are also known, \cf \cite[Sec.~11.7]{Olver}.

The Dirichlet boundary conditions imply the quantization condition
\[ w_{1}(\omega^{+};\hbar) w_{2}(\ell;\hbar)-
w_{2}(\omega^{+};\hbar) w_{1}(\ell;\hbar) =0\,.
\]
Factoring the error in \eqref{eq:WKB1tp sol}
in the classically forbidden region, using the asymptotic expansions of the
Airy functions \cite[p.392/3]{Olver}, and substituting all this into the
quantization condition yields
\[
e^{{\cal S}(\omega^{+})/\hbar}\biggl(
\cos(\frac{{\cal S}(\ell)}{\hbar}-\frac{\pi}{4}) + {\cal O}(\hbar)
\biggr) + \frac{1}{2}
e^{-{\cal S}(\omega^{+})/\hbar}\biggl(
\sin(\frac{{\cal S}(\ell)}{\hbar}-\frac{\pi}{4}) + {\cal O}(\hbar)
\biggr)=0\,.
\]
If this equation is satisfied, then necessarily,
\( \cos(\frac{{\cal S}(\ell)}{\hbar}-\frac{\pi}{4}) = {\cal O}(\hbar)
\).
This implies
\(
        \frac{{\cal S}(\ell)}{\hbar}-\frac{\pi}{4} =
        \frac{2n+1}{2}\pi + {\cal O}(\hbar)
        \),
or equivalently
\[
\int_{x_{t}}^{\ell}\sqrt{E-V(t)}\,dt =
(n+\frac{3}{4})\pi\hbar +{\cal O}(\hbar^{2})\,.
\]
Now using \textbf{(H4)}, we have
\begin{align*}
\int_{x_{t}}^{\ell}\sqrt{E-V(t)}\,dt\,&=\,
\int_{x_{t}}^{\ell}\sqrt{E-v_{+}}\,dt +
\int_{x_{t}}^{\ell}\,(\sqrt{E-V(t)}-\sqrt{E-v_{+}})\,dt
\\
&=\,\sqrt{E-v_{+}}\,(\ell-x_{t})+
\int_{x_{t}}^{\ell}\,\frac{v_{+}-V(t)}{\sqrt{E-V(t)}+\sqrt{E-v_{+}}}\,dt
\\
&=\,\ell\,\sqrt{E-v_{+}}\,(1+{\cal O}(\ell^{-\epsilon}))
\end{align*}
From this, it follows that
\[
E=v_{+}+\left(\frac{(n+3/4)\,\pi\hbar}{\ell}\right)^{2}\left(1+
{\cal O}(\hbar) +{\cal O}(\ell^{-\epsilon})\right)\,.
\]

\begin{thebibliography}{1}

\bibitem{Agmon}
S.~Agmon.
\newblock {\em Lectures on Exponential Decay of Solutions of Second Order
  Elliptic Operators}.
\newblock Number~29 in Princeton Mathematical Notes. Princeton University
  Press, 1982.

\bibitem{AguCom71}
J.~Aguilar and J.-M. Combes.
\newblock A class of analytic perturbations for one-body schr{\"o}dinger
  hamiltonians.
\newblock {\em Commun. Math. Phys.}, 22:269--279, 1971.

\bibitem{BriComDuc87a}
Philippe Briet, Jean-Michel Combes, and Pierre Duclos.
\newblock On the location of resonances for {Schr{\"o}dinger} operators in the
  semiclassical limit {I}: {Resonance} free domains.
\newblock {\em J.Math.Anal.Appl.}, 125:90--99, 1987.

\bibitem{ComDucKleSei87a}
Jean-Michel Combes, Pierre Duclos, Markus Klein, and Ruedi Seiler.
\newblock The shape resonance.
\newblock {\em Commun. Math. Phys.}, 110:215--236, 1987.

\bibitem{Dieudonne}
Jean {Dieudonn\'e}.
\newblock {\em Calcul {Infinit{\'e}simal}}.
\newblock Collection M\'ethodes. Hermann, Paris 1968.

\bibitem{KukulinEtAl}
V.I.~Kukulin \emph{et al}.
\newblock {\em Theory of Resonances, Principles and Applications}.
\newblock Reidl Texts in the Mathematical Sciences. Kluwer Academic Publishers,
  1989.

\bibitem{Kato}
Tosio Kato.
\newblock {\em Pertubation Theory for Linear Operators}.
\newblock Number 132 in Grundlehren der mathematischen Wissenschaften.
  Springer, Berlin, 3rd edition, 1980.

\bibitem{MaiCedDom80a}
C.H. Maier, L.S. Cederbaum, and W.~Domcke.
\newblock A spherical-box approach to resonances.
\newblock {\em J. Phys. B: Atom. Molec. Phys}, 13:L119--L124, 1980.

\bibitem{Olver}
F.~W.~J. Olver.
\newblock {\em Asymptotics and Special Functions}.
\newblock Computer Science and Applied Mathemetics. Academic Press, New York
  and London, 1974.

\bibitem{Reed-SimonIV}
Michael Reed and Barry Simon.
\newblock {\em Analysis of Operators}, volume~IV of {\em Methods of Modern
  Mathematical Physics}.
\newblock Academic Press, San Diego, 1978.

\bibitem{WigVNeu27a}
Eugene~P. Wigner and von Neumann.
\newblock {\"Uber das Verhalten von Eigenwerten bei adiabatischen Prozessen}.
\newblock {\em Zeitschr. f. Phys}, 1927.

\end{thebibliography}

\end{document}